\documentclass[letterpaper]{article}
\usepackage[preprint]{aaai2027}
\usepackage[hyphens]{url}
\usepackage{graphicx}
\usepackage{natbib}
\usepackage{caption}
\usepackage{booktabs}
\usepackage{amsmath}
\usepackage{amssymb}
\usepackage{algorithm}
\usepackage{algorithmic}
\usepackage[table]{xcolor}
\usepackage{tikz}
\usetikzlibrary{arrows.meta,calc,positioning}
\definecolor{LCGreen}{HTML}{006955}
\definecolor{LCBlue}{HTML}{245C8A}
\definecolor{LCOrange}{HTML}{B45214}
\definecolor{LCRed}{HTML}{A02D19}
\newcommand{\good}[1]{\textcolor{LCGreen}{\textbf{#1}}}
\newcommand{\bestours}[1]{\cellcolor{LCGreen!11}\textcolor{LCGreen}{\textbf{#1}}}
\newcommand{\bestpeer}[1]{\cellcolor{LCBlue!10}\textcolor{LCBlue}{\textbf{#1}}}
\newcommand{\failstatus}[1]{\cellcolor{LCRed!7}\textcolor{LCRed}{\textit{#1}}}
\newcommand{\tiecell}[1]{\cellcolor{LCOrange!10}\textcolor{LCOrange}{\textbf{#1}}}
\newcommand{\localstatus}[1]{\cellcolor{black!5}\textcolor{black!65}{\textit{#1}}}
\newcommand{\yes}{\cellcolor{LCGreen!12}\textcolor{LCGreen}{\textbf{Y}}}
\newcommand{\no}{\cellcolor{black!4}N}
\newcommand{\external}{\cellcolor{LCBlue!9}\textcolor{LCBlue}{\textbf{E}}}
\newcommand{\approximate}{\cellcolor{LCOrange!11}\textcolor{LCOrange}{\textbf{A}}}
\newcommand{\unspec}{\cellcolor{black!3}\textcolor{black!45}{--}}
\graphicspath{{figures/}}
\newcommand{\method}{LC-Implicit-QAOA}

\begin{document}

\title{LC-Implicit-QAOA: Active-Workspace-Capped Exact Objective-and-Gradient Evaluation for Training over Bounded QUBO Light Cones}
\author{Chih-Chung Hsu}
\affiliations{
    Institute of Smart Industry and Green Energy,\\
    National Yang Ming Chiao Tung University, Hsinchu, Taiwan\\
    chihchung@nycu.edu.tw\\[3pt]
    {\small Technical supplement and code:
     \url{https://github.com/jesse1029/lc-implicit-qaoa}}
}
\maketitle

\begin{abstract}
QAOA training repeatedly queries an objective and all shared gradients, making exact evaluation a feasibility bottleneck even when QUBO terms have bounded causal cones. Building on established causal-cone restriction and adjoint differentiation, \method{} profiles cone structure and induced-edge counts before local-amplitude and named-workspace allocation, then jointly selects equal-size microbatches and checkpoint schedules under a named active-evaluator workspace budget. ``Implicit'' means omitting both global state and global cost table, not implicit differentiation; infeasible requests are rejected before those allocations. An independently implemented complex128/float64 dense adjoint agrees with LC over 1,800 graph--angle comparisons, with a worst relative gradient error of $1.56\times10^{-13}$. LC completes all 104 target requests in a $p{=}2$ bounded-cone grid; under a prespecified $n\leq24$ validation cap, the matched state-plus-cost reference is executed for 28 requests and deliberately not run on 76. Across 80 budgeted requests, measured allocated evaluator memory stays within budget, reaching at most 0.797 of it. On 3-regular $n{=}512,p{=}2$, the adjoint reaches the same finite-budget endpoint in 101 objective-equivalent calls and 189~s, versus 909 calls and 1,565~s for central differences. LC targets fixed-depth one- and two-local diagonal QUBO costs with a transverse-field mixer; it provides neither global states, sampling, nor a hardware-independent fastest-backend rule.
\end{abstract}

\section{Introduction}

In QUBO-based AI search and hybrid optimization, a QAOA optimizer may issue hundreds of exact value and gradient queries before returning useful angles. The evaluator is therefore part of the optimization loop, and its memory and runtime can determine whether a QUBO instance is practically trainable in an exact simulator. This limitation is not captured by the circuit size alone: when the cost is a sum of local terms, the relevant causal cones can remain moderate even as the global state becomes infeasible.

\begin{figure*}[t]
  \centering
  \includegraphics[width=\textwidth]{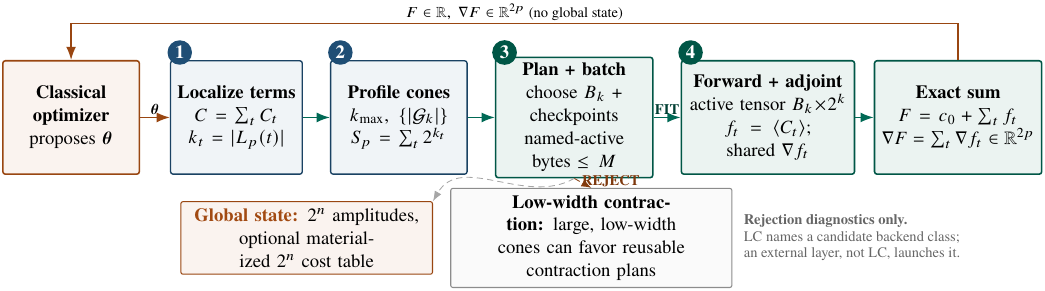}
  \caption{LC inside the optimizer loop. Before local-amplitude or named-workspace allocation, LC profiles cone structure and induced-edge counts, then fits a batch--checkpoint plan under $M$ or rejects the request. A fit returns $F\in\mathbb R$ and $\nabla F\in\mathbb R^{2p}$ without a global state. The lower band lists backend classes as rejection diagnostics; an external layer, not LC, launches them, and the profile does not rank their speed.}
  \label{fig:workflow}
\end{figure*}

QUBO supplies the objectives such loops are trained on, from graph objectives to feature-subset surrogates and NP-hard encodings used in AI search~\cite{glover2018qubo,lucas2014ising,peng2005mrmr,mucke2023fsqubo}, and QAOA turns each into a differentiable model by alternating a diagonal cost with a mixer~\cite{farhi2014qaoa,blekos2024review}. Full-state engines evolve $2^n$ amplitudes and often materialize a $2^n$ cost vector per query. LC materializes neither global object; ``implicit'' refers to the state and cost table, not differentiation. Each term is instead restricted to its causal cone, batched by size, and differentiated by reverse mode. The optimizer, model, and landscape are unchanged; only the state materialized during training changes (Figure~\ref{fig:workflow}).

Causal-cone restriction, adjoint differentiation, and their use for local shared-parameter gradients are established capabilities. LC does not claim these identities as new. Its AI-systems contribution is the execution contract around them: pre-allocation profiling, a joint byte model for heterogeneous batching and checkpoint selection, and fit-or-reject execution inside QUBO optimization loops.

The pre-allocation profile $\Pi_p(G)=(k_{\max},\{|\mathcal G_k|\}_k,S_p)$, where $S_p=\sum_t2^{k_t}$, controls cone occupancy and memory feasibility; complete operation work also depends on $|E[L_p(t)]|$. An instance is \emph{profile-feasible} when these quantities fit the dense-local budget. Bounded-degree and degree-capped constructions often are; sparsity alone is insufficient.

The paper makes two contributions.
\begin{itemize}
    \item \textbf{A profile-and-plan execution contract for exact cone-local QUBO-QAOA queries.} Given a declared named active-evaluator workspace budget $M$, the planner jointly selects equal-size microbatches and checkpoint subsets, or rejects the request before local-amplitude and named-workspace allocation when no plan satisfies the contract.
    \item \textbf{A reproducible characterization of the regime in which this contract is useful.} The evaluation separately tests algebraic and numerical exactness, budget compliance, bounded-cone reach, repeated optimizer use, and backend-specific runtime, while retaining dense, hub-driven, and low-width contraction regimes as explicit limitations.
\end{itemize}

The implementation covers fixed-depth QAOA with one- and two-local diagonal costs and the transverse-field mixer. Real-data-derived QUBOs test its workload structure rather than a new downstream predictor, and global state access and final sampling belong to a separate backend.

\section{Related Work}

\textbf{Local evaluation of QAOA.}
At fixed depth, each term depends only on its distance-$p$ neighborhood~\cite{farhi2014qaoa,streif2020training}. This supports causal-cone tensor training~\cite{streif2020training}, fixed-angle reuse of isomorphic regular-graph environments~\cite{wurtz2021subgraphs}, QTensor's local expectation path~\cite{lykov2022qtensor}, $p{=}1$ closed forms~\cite{ozaeta2022expectation}, and correlator light cones for depth-two expectations on nonplanar regular graphs~\cite{sack2024large}. The closest implementation is Qcover~\cite{zhuang2021qcover}, which decomposes weighted node and edge terms into $p$-hop neighborhoods and sums their expectations into the exact global objective. We verified its official release (supplement): it returns a scalar objective and leaves derivatives to numerical or derivative-free optimizers. Weighted decomposition is therefore prior work, and the boundary lies in the gradient and execution contract. Locality also underlies QAOA approximation limits~\cite{hastings2019boundeddepth,marwaha2021localmaxcut,basso2022constantqaoa}, and Wang et al.~\cite{wang2025constantqaoa} parameterize constant-round evaluation by $p$-local treewidth, complementing our cone-cardinality route.

\textbf{Differentiable tensor-network frameworks.}
TensorCircuit and its maintained successor TensorCircuit-NG both expose light-cone-reduced local expectations under automatic differentiation, so exact local objectives with all shared parameter derivatives are prior capability~\cite{zhang2023tensorcircuit,zhang2026tcng}. TensorCircuit-NG also performs reusable pathfinding and slicing through a differentiable sliced contractor. We verified both official releases directly (supplement). Its slicing target bounds the largest intermediate tensor of one contraction, not the simultaneously active tensors of a complete objective-and-gradient execution. Neither release exposes profiled equal-$k$ QUBO-cone batching, byte-model checkpoint selection, or fit-or-reject before local-amplitude and named-workspace allocation as a native primitive. These frameworks are the closest complementary route, not a baseline LC supersedes; large low-width cones remain their regime.

\textbf{Global-state execution.}
QOKit precomputes the diagonal Hamiltonian~\cite{lykov2023qokit}; CUAOA provides CUDA-native expectations, gradients, states, and samples~\cite{stein2024cuaoa}; CUDA-Q, cuStateVec, PennyLane-Lightning-GPU, and JuliQAOA provide global-state execution~\cite{nvidia2026cudaq,bayraktar2023cuquantum,asadi2024lightning,golden2023juliqaoa}; MPS-JuliQAOA and BMQSim trade exactness for bond dimension or amplitude compression~\cite{feeney2025mpsjuliqaoa,zhang2025bmqsim}. LC keeps the constant-pass adjoint of Jones and Gacon~\cite{jones2020gradients} but replaces one global state with a stream of cone-local batches, so its cost is governed by cone cardinality and occupancy where full-state adjoints are governed by $n$.

\begin{table}[t]
\centering
\small
\setlength{\tabcolsep}{1.0pt}
\begin{tabular}{@{}lccc@{\hskip 6pt}cccc@{}}
\toprule
& \multicolumn{3}{c}{Established capability} & \multicolumn{4}{c}{Execution primitive}\\
\cmidrule(r){2-4}\cmidrule(l){5-8}
Method & Local & Grad & Shared & Eq.-$k$ & N-byte & B-ckpt & Pre-fit\\
\midrule
\rowcolor{LCGreen!7}\textcolor{LCGreen}{\textbf{LC (ours)}} & \yes & \yes & \yes & \yes & \yes & \yes & \yes\\
Qcover & \yes & \external & \external & \no & \no & \no & \no\\
TensorCircuit & \yes & \yes & \yes & \no & \no & \no & \no\\
TC-NG & \yes & \yes & \yes & \no & \no & \no & \no\\
\midrule
QTensor & \yes & \external & \external & \unspec & \unspec & \unspec & \unspec\\
QOKit & \no & \external & \external & \unspec & \unspec & \unspec & \unspec\\
CUAOA & \no & \yes & \yes & \unspec & \unspec & \unspec & \unspec\\
CUDA-Q/cuSV & \no & \external & \external & \unspec & \unspec & \unspec & \unspec\\
PL-Lightning & \no & \yes & \yes & \unspec & \unspec & \unspec & \unspec\\
MPS-JQAOA & \no & \approximate & \approximate & \unspec & \unspec & \unspec & \unspec\\
\bottomrule
\end{tabular}
\caption{Established local-differentiation capabilities and the narrower execution primitives compared here. Eq.-$k$, N-byte, B-ckpt, and Pre-fit denote profiled equal-$k$ cone batching, a named-active byte model, byte-model checkpoint selection, and pre-amplitude fit-or-reject. Y is a verified native exact interface, E an external numerical route, A approximate support, and N no such native primitive in the cited release; N does not preclude assembly from lower-level pieces. -- marks unverified cells.}
\label{tab:capability}
\end{table}

\section{Problem Setting}

Let $G=(V,E)$ be the interaction graph of a QUBO instance with $|V|=n$. Binary variables $x_i\in\{0,1\}$ map to Ising variables $z_i=1-2x_i\in\{-1,+1\}$, after which the diagonal objective is
\begin{equation}
  C(z) = c_0 + \sum_{i \in V} h_i z_i + \sum_{(i,j)\in E} J_{ij} z_i z_j .
  \label{eq:qubo}
\end{equation}
The evaluator stores $c_0$ with the graph and returns it inside objective values; it affects neither state evolution, gradients, nor supports, and MaxCut is the special case with edge terms plus an offset. A depth-$p$ QAOA state is
\begin{equation}
  |\psi(\boldsymbol{\gamma},\boldsymbol{\beta})\rangle
  = \prod_{\ell=1}^{p} e^{-i\beta_\ell B} e^{-i\gamma_\ell C}|+\rangle^{\otimes n},
  \qquad B=\sum_i X_i,
  \label{eq:state}
\end{equation}
and the training objective is $F(\boldsymbol{\gamma},\boldsymbol{\beta})=\langle\psi|C|\psi\rangle$. The full-state route stores $|\psi\rangle\in\mathbb{C}^{2^n}$ and often a $2^n$ diagonal table for $C(z)$. \method{} evaluates each term's expectation on its light cone instead. Throughout, \emph{exact} denotes equality to the full QAOA objective and gradient in exact arithmetic; reported discrepancies arise from finite floating-point precision or from a separately identified comparator adapter.

\section{Method}

\subsection{Light-Cone Extraction}

Write $C=\sum_{t\in T}C_t$ with each nonconstant term supported on a node or edge set $S_t$, so that $\langle C\rangle=\sum_t\langle C_t\rangle$ exactly. For a term $t$, define $L_p(t)$ as the set of qubits reachable from $S_t$ by at most $p$ graph hops and $k_t=|L_p(t)|$. Gates outside $L_p(t)$ cancel when evaluating $\langle C_t\rangle$, so
\begin{equation}
  \langle C_t\rangle_G =
  \langle +|_{L_p(t)}
  U_{L_p(t)}^\dagger C_t U_{L_p(t)}
  |+\rangle_{L_p(t)} ,
  \label{eq:local}
\end{equation}
where $U_{L_p(t)}$ is the QAOA unitary restricted to the induced weighted subgraph $G[L_p(t)]$ with its couplings, fields, and angles. The cost of evaluating \eqref{eq:local} is exponential in $k_t$, not in $n$.

\textbf{Proposition 1 (standard causal-cone identity; restated).}
For the one- and two-local diagonal Hamiltonian and transverse-field mixer $B=\sum_iX_i$ implemented here, the expectation of each nonconstant term $C_t$ after $p$ layers equals its expectation under the QAOA circuit of the induced weighted subgraph on $L_p(t)$. This known identity~\cite{farhi2014qaoa,streif2020training,wurtz2021subgraphs,lykov2022qtensor} is restated because the implementation must preserve arbitrary edge weights and linear fields.

\textbf{Proof sketch.}
In the Heisenberg picture, one-qubit mixer factors do not move support between qubits, and a two-body cost factor can enlarge support only across an incident edge. Induction over $p$ backward cost layers confines support to the closed $p$-hop neighborhood $L_p(t)$; every surviving factor lies in the induced subgraph with its original weight, and $|+\rangle^{\otimes n}$ factorizes across the boundary. The supplement formalizes the boundary and restriction steps.

\subsection{Batched Local Simulation and Adjoint}

Figure~\ref{fig:workflow} and Algorithm~\ref{alg:lc-evaluator} give the execution path. For $\mathcal{G}_k=\{t:k_t=k\}$, chunks form $(B_k,2^k)$ tensors, run the local circuit, and reverse its layers into the shared $2p$ slots before release. All buckets are planned before local-amplitude and named-workspace allocation; an infeasible bucket returns a rejection with diagnostics, while a feasible request executes one chunk at a time and forms no global array.

\noindent\textbf{Planner specification.}
For each nonempty $\mathcal G_k$, candidates are $1\leq B_k\leq\min(|\mathcal G_k|,\lfloor B_{\rm state}/2^k\rfloor)$ and every $\mathcal C_k\subseteq\{0,\ldots,p-2\}$. Predicted active bytes $A_k$ sum cost/term tables, final, adjoint, checkpoint, reconstruction, derivative, mixer, observable, and gradient workspaces. With $q_k=\lceil|\mathcal G_k|/B_k\rceil$, feasible candidates satisfy $A_k\leq M$ and lexicographically minimize $(q_k[p+R(\mathcal C_k)],q_k,-B_k,A_k)$, where $R$ counts reconstructed layer steps; this estimates work, not wall time. One chunk is resident; $M$ covers only $A_k$ and excludes cone metadata, allocator reserve, CUDA context, and whole-process/device memory. If any bucket has no candidate, LC rejects before local-amplitude and named-workspace allocation and reports, but does not launch, alternative backend classes.

\begin{algorithm}[t]
\caption{Named-workspace-capped objective and shared-gradient evaluation}
\label{alg:lc-evaluator}
\small
\begin{algorithmic}[1]
\REQUIRE QUBO $G$, depth $p$, angles $\theta$, named workspace budget $M$
\STATE extract $\{L_p(t)\}$ and form profile $\Pi_p(G)$; $F\leftarrow c_0$, $g\leftarrow0_{2p}$
\FOR{each size bucket $\mathcal G_k$}
  \STATE $(B_k,\mathcal C_k)\leftarrow\textsc{Plan}(k,|\mathcal G_k|,p,M,\mathrm{dtype})$
  \IF{infeasible} \RETURN $\textsc{Reject}(\Pi_p(G),M,\text{diagnostics})$ \ENDIF
\ENDFOR
\FOR{each planned size bucket $\mathcal G_k$}
  \FOR{each chunk $A\subseteq\mathcal G_k$, $|A|\leq B_k$}
    \STATE $(f_A,g_A)\leftarrow\textsc{LocalAdjoint}(A,\theta,\mathcal C_k)$; $(F,g)\mathrel{+}=(\sum f_A,\sum g_A)$
  \ENDFOR
\ENDFOR
\RETURN $(F,g,\Pi_p(G))$
\end{algorithmic}
\end{algorithm}

Where central finite differences need $4p$ objective evaluations, the reverse pass prices the whole gradient at a constant factor over one forward pass: for a parameterized local unitary $U_\alpha=e^{-i\alpha H}$,
\begin{equation}
  \frac{\partial \langle C_t\rangle}{\partial \alpha}
  = 2\,\mathrm{Re}\left\langle \lambda_{\mathrm{after}},
  (-iH)\, U_\alpha \psi_{\mathrm{before}}\right\rangle ,
  \label{eq:adjoint}
\end{equation}
with $\langle a,b\rangle=a^\dagger b$, exactly the adjoint scheme used for full-state simulated circuits~\cite{jones2020gradients}, applied here per cone with cross-cone accumulation into the shared angles. Restricting the mixer sum to the cone is valid because identity \eqref{eq:local} holds for all angles, so derivative contributions of outside factors vanish identically.

\textbf{Corollary 1 (cone-local shared-parameter adjoint).}
In exact arithmetic and under Proposition~1's assumptions, the local forward--reverse procedure returns $\nabla_\theta F$ for the same objective as global-state QAOA: differentiate Eq.~\eqref{eq:local} term by term and sum the shared-parameter derivatives. Numerical discrepancies can therefore arise only from finite precision or from a numerical reference.

\subsection{Four Memory Quantities}
\label{sec:memory-terms}

We distinguish four quantities: \emph{analytical scaling bytes} ($B2^k$), \emph{planner-predicted active bytes} for the named active-evaluator workspaces, \emph{measured evaluator allocation}, and separate \emph{allocator-reserved or process/device peak memory}. Every cap statement identifies its numerator and denominator. The budget $M$ applies to predicted active bytes; Section~\ref{sec:budget} compares measured allocation with $M$ and the prediction.

\subsection{Pre-allocation Profile and Bounds}
\label{sec:bounds}

The state-plus-cost route stores $\Theta(2^n)$ complex amplitudes plus a $\Theta(2^n)$ real cost table and reads the global distribution per query. Cone extraction instead yields $\Pi_p(G)$ before local-amplitude and named-workspace allocation; together with the selected $B_k$, checkpoint multiplier $m_k$, and recomputation factor $r_t\geq1$, it fixes cone-size occupancy, chunk counts, and predicted active bytes. Complete operation work additionally depends on each cone's induced-edge count, as expressed by
\begin{equation}
  W_{\mathrm{LC}} =
  \sum_{t\in T} O\!\left(r_t p\,(|E[L_p(t)]|+k_t)\,2^{k_t}\right),
  \label{eq:work}
\end{equation}
with analytical scaling bytes $M_{\mathrm{obj}}=O(\max_k B_k2^k)$ and $M_{\mathrm{grad}}=O(\max_k B_km_k2^k)$. Here $m_k\geq1$ is the retained state and workspace multiplier induced by the selected reverse schedule, and $r_t$ is the forward-reconstruction factor it implies; cache-all has $r_t=1$ and the largest $m_k$. Both describe the schedule the planner selects under $M$, not a proof that the selection is globally optimal. The exponent is $k$, not $n$.

\textbf{Proposition 2 (parameterized work and active-bucket residency).}
Under Proposition~1, the batched local procedure has work given by Eq.~\eqref{eq:work}, up to batch-launch overhead, and never holds more than $O(B_k2^k)$ objective or $O(B_km_k2^k)$ adjoint amplitudes for the bucket being executed. The statement bounds active-bucket residency under the byte model of Section~\ref{sec:memory-terms}, not the total process footprint, and chunk count and $r_t$ affect time rather than that bound.

\textbf{Corollary 2 (bounded-degree fixed-depth regime).}
If $|T|=O(n)$, supports are bounded, the maximum degree is $\Delta=O(1)$, and $p=O(1)$, then $k_t=O(\Delta^p)$ and the implemented checkpoint schedules have $r_t=O(1)$, giving $W_{\mathrm{LC}}=O(n\,2^{O(\Delta^p)})$: linear in $n$ with the constant set by the cone. Edge sparsity alone is insufficient: a star has $n-1$ edges but a hub-edge cone of size $n$ at $p=1$.

\section{Experimental Setup}

\textbf{Hardware.}
The cold-objective matrix, 8~GiB boundary, and checkpoint-planner study use an RTX~3070, with the CUAOA comparison replicated on an RTX~3090. The $p{=}2$ family grid and the microbatch sweep use an RTX~3090, and the dense-local/QTensor grid matched cells on RTX~3090/4090 GPUs. Absolute times are not compared across hosts, and every reported ratio is formed within a device- and precision-matched protocol. Backend versions, the documented CUAOA device-information patch, adapter diagnostics, and build details appear in the supplement.

\begin{figure*}[t]
  \centering
  \includegraphics[width=\textwidth]{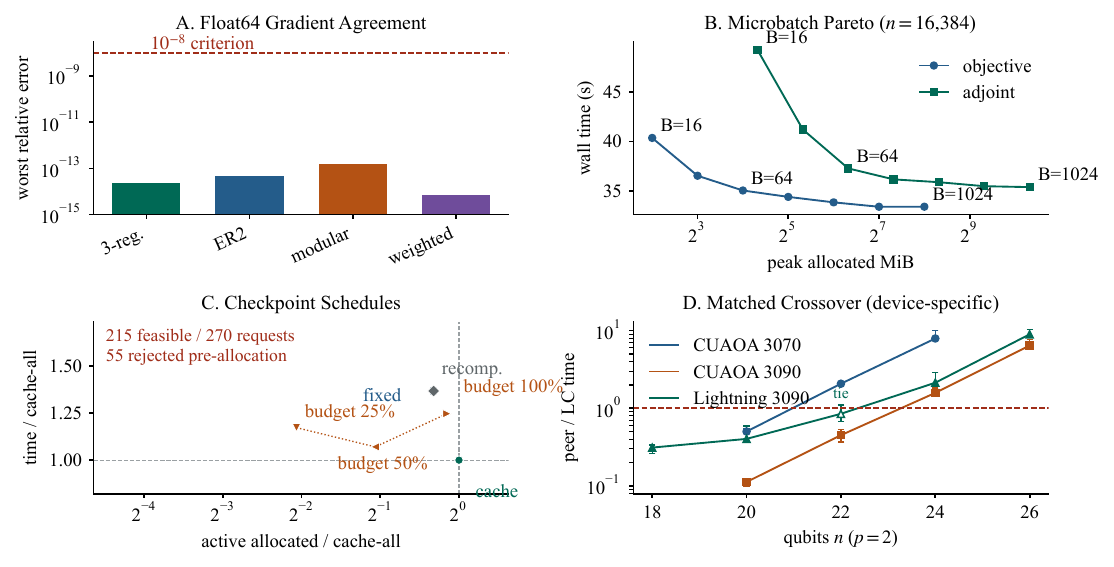}
  \caption{Evidence in claim order. (A) Independent float64 correctness over 1,800 graph--angle pairs. (B) Measured allocated memory follows $B2^k$ while runtime saturates by $B{=}64$. (C) Four schedule families; budget labels give $M$/cache-all and the horizontal axis is allocated evaluator memory. Pool-reserved and device peaks are distinct and unplotted. (D) Same-device, matched-precision $p{=}2$ global-adjoint comparisons; \emph{tie} means the paired 95\% interval contains peer/LC $=1$.}
  \label{fig:evidence}
\end{figure*}

\textbf{Measurement protocol.}
Timings are wall-clock, and allocated and allocator-reserved memory are reported independently. We separate preprocessing, the first query, and post-warmup steady queries: Table~\ref{tab:peer} includes cone extraction and first compilation, while Figure~\ref{fig:evidence}D uses post-warmup medians over five repeats. The $p{\geq}2$ sweep uses ten graph seeds per family-size cell with structural eligibility fixed before timing, and the float64 study uses ten graphs and five angle initializations per family-size-depth cell.

\textbf{Resource limits.}
The five quantities below are not interchangeable.
\begin{center}
\scriptsize
\setlength{\tabcolsep}{2.0pt}
\renewcommand{\arraystretch}{0.90}
\begin{tabular}{@{}p{.30\columnwidth}p{.64\columnwidth}@{}}
\toprule
Name & Definition and use\\
\midrule
Analysis target & $k_{\max}\leq14,\ S_p\leq2{\times}10^7$; selects the primary $p{=}2$ grid.\\
Per-cone guardrail & $k\leq24$; bounds each dense local state.\\
Experiment cone-state volume cap & $S_p\leq2^{30}$ in sweeps; $2^{31}$ in extended reach.\\
State+cost protocol cap & $n\leq24$ in the family grid; larger rows are non-runs.\\
Measured boundary & RTX~3070 c64/f32 state+cost: $n=26$ succeeds, $n=28$ OOM.\\
\bottomrule
\end{tabular}
\end{center}

\textbf{Graph families and comparators.}
We study 3-regular graphs, sparse Erd\H{o}s--R\'enyi (ER) graphs with expected degree 2 to 4, modular sparse QUBOs, weighted QUBOs with fields, dense ER graphs, and scale-free graphs; the last two separate controlled local growth from edge sparsity. The matched CuPy references are a global state plus precomputed diagonal table and a dense global-state adjoint, and the external systems are CUAOA, PennyLane-Lightning-GPU, CUDA-Q, QTensor CPU/GPU, and QOKit CPU/GPU. Runtime comparisons use only the named upstream path, and adapter crashes, timeouts, and unsupported outputs stay separate statuses rather than performance losses. Generators, seeds, per-seed intervals, and status definitions are given in the supplement.

\section{Results}

\subsection{Does LC return the same objective and all shared gradients?}

Yes, to the precision of an independent implementation. A separately implemented dense global-state adjoint in complex128/float64 is compared with LC for four families, $n\in\{8,12,16\}$, $p\in\{1,2,3\}$, ten graphs, and five angle seeds: 1,800 graph--angle comparisons. The worst relative $\ell_2$ error is $1.56\times10^{-13}$, the worst component error $3.22\times10^{-13}$, and the minimum cosine similarity differs from one by less than $5\times10^{-16}$ (Figure~\ref{fig:evidence}A). Cell-level median directional Taylor slopes are approximately two. Central differences serve only as a secondary reference, since their own truncation error exceeds the discrepancy under test. Exactness is an algebraic property of the decomposition; this experiment bounds the numerical realization of it, not the runtime behavior discussed later.

\subsection{Does the declared budget hold, and which requests become feasible?}
\label{sec:budget}

The checkpoint-planner study issues $3\ \text{settings}\times3\ \text{graph instances}\times6\ \text{configurations}\times5\ \text{process repetitions}=270$ requests. The six configurations are \textsc{cache-all}, \textsc{recompute-all}, \textsc{fixed-interval}, and budgeted at 25\%, 50\%, and 100\% of the cache-all prediction; 215 requests complete and 55 are rejected before local-amplitude and named-workspace allocation, which together account for all 270. The 80 completed budgeted requests are the ones that carry a declared $M$. In every one of them, predicted active bytes stay within $M$, reaching at most 0.974 of it, and \emph{measured} allocated evaluator memory also stays within $M$, reaching at most 0.797 of it. Across all 215 feasible requests, measured allocation never exceeds the planner prediction, with a maximum measured-to-predicted ratio of 0.857. Allocator-reserved memory exceeds $M$ in all 80 budgeted requests, as expected for a retained CuPy pool; reservation is reported as a separate observation and is not the quantity the planner bounds. Accuracy is unchanged: relative gradient error over these runs is at most $3.44\times10^{-8}$ in complex64 and $2.59\times10^{-15}$ in complex128.

Two prediction models appear in this paper and must not be quoted as one. The 1.053 ratio compares measured allocation with the simplified $B2^k$ estimate of Section~\ref{sec:bounds}, which predicts 76 bytes per active local amplitude against a measured 80, and is a diagnosis of that estimate rather than a cap-compliance test. The 0.857 ratio instead compares measured allocation with the complete scheduling byte model, which conservatively sums named workspaces that do not all coexist at the high-water mark.

\begin{table}[t]
\centering
\small
\setlength{\tabcolsep}{2.5pt}
\begin{tabular}{@{}llrrrr@{}}
\toprule
Case & Schedule & $M$ & Alloc. & Pool & \multicolumn{1}{c}{s}\\
& & \% & MiB & MiB & sel./cache\\
\midrule
3-reg. $512,p2$ & budgeted & 25 & \good{19.1} & 41.3 & 1.024/.875\\
3-reg. $24,p3$ & budgeted & 100 & \good{640} & 1,351 & 3.856/2.952\\
wER2 $96,p2$ & budgeted & 100 & \good{576} & 1,284 & 3.027/2.428\\
\bottomrule
\end{tabular}
\caption{Non-dominated checkpoint plans: RTX~3070 c64/f32 steady value-plus-gradient. Values are medians over three graphs after five process repeats each. Alloc.\ is tested against $M$; Pool is reserved and outside the contract. Whole-device peak is distinct and unavailable for these rows.}
\label{tab:budgeted}
\end{table}

Within that budget the planner is what makes cone-local execution fast. Equal-size batching gives $6.11\times$ speedup at 3-regular $n{=}24,p{=}2$, $3.54\times$ at $n{=}512,p{=}2$, and $4.91\times$ on modular weighted QUBO $n{=}128,p{=}1$ over one-cone execution, without changing the objective, and the selected checkpoint subsets differ from fixed intervals (Table~\ref{tab:budgeted}). Measured allocated memory follows $B2^k$: at $n{=}16{,}384,p{=}2,k_{\max}{=}14$, raising $B$ from 16 to 1,024 lifts objective and adjoint allocation from 4.0/20.0 to 256/1,280~MiB while time improves only 17/28\%, and at $B{=}64$ allocation stays at 16/80~MiB for $n{=}512$ through 16,384 (Figure~\ref{fig:evidence}B).

Feasibility follows the same profile. Inside the analysis target band, LC completes all 104 instance-level value-plus-gradient requests across four families, while the matched state-plus-cost reference completes 28 (Table~\ref{tab:p2}). Its 76 remaining instances are declared non-runs under the prespecified protocol cap, not observed OOMs. A near-cap band adds 83 further completions, whereas dense ER and scale-free families exceed the same local-state cap in most settings and complete only 10/60 each (Table~\ref{tab:boundary}). Those failures stay in the regime matrix but are excluded from the runtime regression, which is defined only over completed LC queries.

\begin{table}[t]
\centering
\small
\setlength{\tabcolsep}{2.2pt}
\begin{tabular}{@{}lrrrrr@{}}
\toprule
Family & $n$ & \shortstack{LC\\done} & Obj./grad. s & LC MiB & \shortstack{Global\\done}\\
\midrule
\rowcolor{LCGreen!5}3-regular & 24--512 & \good{80/80} & .126/.150 & 72.4 & 10/80\\
\rowcolor{LCGreen!5}ER2 & 24--32 & \good{7/7} & .0275/.0432 & 2.01 & 3/7\\
\rowcolor{LCGreen!5}Modular & 24--64 & \good{11/11} & .0149/.0201 & .021 & 10/11\\
\rowcolor{LCGreen!5}Weighted & 24--64 & \good{6/6} & .0619/.0922 & 4.75 & 5/6\\
\midrule
\textbf{Total} & 24--512 & \bestours{104/104} & -- & -- & \textbf{28/104}\\
\bottomrule
\end{tabular}
\caption{Family-separated $p{=}2$ grid in the analysis target band ($k_{\max}\leq14$, $S_p\leq2\times10^7$), run on RTX~3090. Each family--size generator cell requests up to ten seeds; totals are structurally eligible, deduplicated graph instances. Times are objective and value-plus-gradient medians; LC MiB is median allocated evaluator memory. LC completes all 104 targets. Under the prespecified $n\leq24$ validation cap, the state-plus-cost reference is executed for 28 requests and deliberately not run on 76; these are protocol-cap non-runs, not OOMs.}
\label{tab:p2}
\end{table}

\begin{table}[t]
\centering
\small
\setlength{\tabcolsep}{4.2pt}
\begin{tabular}{@{}lrrl@{}}
\toprule
Cone family & Requested & Complete & Regime\\
\midrule
\rowcolor{LCGreen!5}ER2 near-cap & 39 & \good{39} & bounded\\
\rowcolor{LCGreen!5}Modular near-cap & 18 & \good{18} & bounded\\
\rowcolor{LCGreen!5}Weighted near-cap & 26 & \good{26} & bounded\\
\rowcolor{LCOrange!6}Dense ER & 60 & 10 & expanding\\
\rowcolor{LCOrange!6}Scale-free & 60 & 10 & hub-driven\\
\bottomrule
\end{tabular}
\caption{Completion counts under a common dense-local cap. The independent unit is one requested graph instance, and no runtime enters the regime label.}
\label{tab:boundary}
\end{table}

\subsection{What does the evaluator change inside an optimizer loop?}

Repeated exact queries are the setting the contract is meant for, so we exercise the adjoint inside Adam (Table~\ref{tab:optimizer}). For 100 updates from the same initialization at $n{=}512,p{=}2$, the shared adjoint uses 101 objective-equivalent evaluator invocations and 189~s, versus 909 invocations and 1,565~s for central differences; the finite-budget endpoints agree to $1.1\times10^{-6}$ relative objective difference. This is neither a global-optimum claim nor a convergence certificate. At $n{=}24$ the full-state finite-difference route reaches the same finite-budget endpoint at 832~MiB against 32.8~MiB for the cone-local adjoint, and it does so in $8.3\times$ the wall time. The optimizer is not a contribution here: Adam paired against equal-budget SPSA gives median normalized-improvement differences of 0.0874--0.0957 across 3-regular $n{=}128$--512, establishing only that the adjoint can drive optimization.

\begin{table}[t]
\centering
\small
\setlength{\tabcolsep}{2.6pt}
\begin{tabular}{@{}llrrrr@{}}
\toprule
Case & Gradient route & Calls & Wall s & Final $F$ & MiB\\
\midrule
\rowcolor{LCGreen!5}3-reg.\ $24,p2$ & LC adjoint & \good{101} & \good{69.9} & 26.97838 & 32.8\\
3-reg.\ $24,p2$ & LC finite diff. & 909 & 135 & 26.97841 & 11.0\\
3-reg.\ $24,p2$ & full-state fin.\ diff. & 909 & 581 & 26.97837 & \failstatus{832}\\
\rowcolor{LCGreen!5}3-reg.\ $512,p2$ & LC adjoint & \good{101} & \good{189} & 580.4453 & 153\\
3-reg.\ $512,p2$ & LC finite diff. & 909 & 1565 & 580.4449 & 51.2\\
\bottomrule
\end{tabular}
\caption{Repeated-query consequence in one bounded-cone optimization setting, with one graph instance per case. Calls are objective-equivalent evaluator invocations, wall is total optimization seconds, and MiB is peak measured allocated evaluator memory. The optimizer is not a contribution; the comparison shows how the evaluator changes the cost of reaching the same endpoint.}
\label{tab:optimizer}
\end{table}

The same interface accepts weighted signed couplings and linear fields, because they do not alter support. Real-data-derived QUBOs test whether such workloads enter the profile-feasible regime at all (Table~\ref{tab:realdata}); the experiment claims no feature-selection accuracy, and supervised selectors stay ahead on classifier metrics (supplement). Dense cardinality penalties instead create all-to-all edges and lose this locality.

\begin{table}[t]
\centering
\small
\setlength{\tabcolsep}{3.0pt}
\begin{tabular}{@{}llrrlrr@{}}
\toprule
Dataset & Variant & Cand. & Deg. & Profile & Obj. s & Grad. s\\
\midrule
WDBC & top30 & 30 & 3 & feasible & 0.452 & 0.582\\
MiceProtein & top64 & 64 & 2 & feasible & 0.449 & 0.532\\
\bottomrule
\end{tabular}
\caption{Degree-capped real-data-derived weighted QUBOs at $p{=}2$ on RTX~3070, complex64/float32. Cand.\ is the candidate feature dimension and Deg.\ the construction degree cap. The rows test whether a real-data-derived weighted QUBO enters the profile-feasible regime; they do not claim improved feature-selection accuracy.}
\label{tab:realdata}
\end{table}

\subsection{When is dense cone-local execution the right route?}

On the named devices, precisions, graph families, and query phases, LC becomes competitive beyond the smallest tested sizes. This is an implementation-level regime observation, not a hardware-independent crossover or a universal backend ranking. In cold objective queries (Table~\ref{tab:peer}), CUAOA leads 3-regular $n{=}24$ at 0.099 against 0.159~s while LC reaches 0.042~s after warm-up; at $n{=}26$ LC needs 0.041~s and 8.2~MiB against 1.47~s and 3.3~GiB for the state-plus-cost route, and at $n{=}28$ that route exhausts the 8~GiB device while LC retains $k_{\max}{=}14$. Cost-table-free CUDA-Q still completes at 2.6~GiB, which locates the wall in the tested state-plus-cost implementation rather than in global execution as a class. ER degree-3 at $n{=}24$ gives the complementary result: a $k_{\max}{=}23$ cone makes LC slower and heavier than global kernels.

\begin{table*}[t]
\centering
\small
\begin{tabular}{@{}lrrrrrr@{}}
\toprule
Case $(n,p,k_{\max})$ & LC (ours) & State+cost & CUAOA & CUDA-Q & QTensor & QOKit CPU\\
 & s / pool MiB & s / pool MiB & s & s / device peak MiB & s & s\\
\midrule
\rowcolor{LCBlue!4}3-regular (24, 2, 14) & 0.159 / 14.9 & 0.341 / 832 & \bestpeer{0.099} & 0.185 & 0.213 & 161\\
\rowcolor{LCGreen!4}3-regular (26, 2, 14) & \bestours{0.041 / 8.20} & 1.47 / 3330 & 0.396 & 0.261 & 0.222 & --\\
\rowcolor{LCGreen!4}3-regular (28, 2, 14) & \bestours{0.045 / 12.0} & \failstatus{OOM} & \localstatus{local $-11$} & 1.26 / \good{2662} & 0.238 & --\\
\rowcolor{LCGreen!4}3-regular (128, 2, 14) & \bestours{0.157 / 82.3} & -- & -- & -- & 1.16 & --\\
\rowcolor{LCGreen!4}ER deg.\ 2 (128, 2, 22) & \bestours{0.267 / 354} & -- & -- & -- & 0.637 & --\\
\rowcolor{LCOrange!6}ER deg.\ 3 (24, 2, 23) & 0.433 / 749 & 0.355 / 832 & \bestpeer{0.100} & 0.184 & 0.324 & 159\\
\bottomrule
\end{tabular}
\caption{Cold objective queries: RTX~3070, complex64/float32, one first query per cell including cone extraction and compilation. The independent unit is one graph instance, and each row is an illustrative matched instance rather than a graph-averaged boundary. The memory metric differs by column and is named in the header. Bold marks the fastest completed value per row, so no ranking rests on shading alone; green rows show bounded-cone growth and the orange row a large cone. The CUAOA $n{=}28$ entry is a local adapter return code, not a method limitation, and QTensor uses its faster CPU path.}
\label{tab:peer}
\end{table*}

Matched steady queries locate the same transition (Table~\ref{tab:direct} and Figure~\ref{fig:evidence}D): global adjoints lead at the smallest tested sizes, while LC wins the largest common size in these device- and precision-matched blocks. A median crossing is the first tested $n$ with median peer/LC ratio above one; an LC win additionally requires the paired 95\% interval above one, and three of six crossings remain ties. Backend versions, numerical checks, and setup/cold/steady timings are in the supplement; the independent dense CuPy adjoint, not these timings, remains the correctness reference.

\begin{table}[t]
\centering
\small
\begin{tabular}{@{}llrrlr@{}}
\toprule
Peer/GPU & Family & Cross. & $n^*$ & LC/peer s & Ratio\\
\midrule
\rowcolor{LCGreen!4}CUAOA/3070 & 3-reg. & 22 & 26 & .0649/2.037 & \bestours{31.4}\\
CUAOA/3070 & wER2 & \tiecell{22$^\dagger$} & 26 & .0840/2.033 & \bestours{24.2}\\
\rowcolor{LCGreen!4}CUAOA/3090 & 3-reg. & 24 & 26 & .183/1.176 & \bestours{6.43}\\
CUAOA/3090 & wER2 & \tiecell{24$^\dagger$} & 26 & .228/1.176 & \bestours{5.15}\\
\rowcolor{LCGreen!4}Lightn./3090 & 3-reg. & 24 & 26 & .167/1.495 & \bestours{8.97}\\
Lightn./3090 & wER2 & \tiecell{22$^\dagger$} & 24 & .202/.643 & \bestours{3.18}\\
\bottomrule
\end{tabular}
\caption{Matched $p{=}2$ steady value-plus-gradient queries on the GPU named per row: CUAOA in its native complex128/float64, Lightning in matched complex64/float32. The independent unit is one graph instance; each cell contains five paired graph seeds evaluated by both methods. Cross.\ is the first tested $n$ with median peer/LC $>1$; the $\dagger$ mark and the orange cell both denote a paired 95\% interval containing one, that is, a tie. $n^*$ is the largest common size and Ratio is peer/LC at $n^*$ from unrounded medians.}
\label{tab:direct}
\end{table}

Contraction remains the better route for some large low-width cones. In the repeated-query study with reusable QTensor plans, LC stays faster inside the tested dense-local band, while QTensor completes two low-width cones rejected by the LC allocation guardrail. Held-out errors and cross-host crossover changes rule out a calibrated fastest-backend predictor. The profile is instead a feasibility screen: rejection names a candidate global or contraction route, whose width and runtime an external orchestration layer must assess and launch. One-shot timing definitions, plan-reuse cells, cross-adapter discrepancies, and held-out results are reported in the supplement.

\section{Discussion and Limitations}

LC applies to fixed-depth QAOA with one- and two-local diagonal QUBO costs and a transverse-field mixer. Its structural limit is cone growth: dense penalties, hubs, and larger depth enlarge the local states, while large low-width cones may favor reusable tensor contraction. The reported budget concerns the named active-evaluator workspace, and separate engineering limits, chiefly unfused CuPy kernels and size-only batching, can move the measured crossover but not the cone bound. LC changes the cost and feasibility of exact queries, not the QAOA landscape, downstream predictor quality, or hardware sampling performance.

QUBO formulation therefore affects evaluator cost as well as optimization semantics: degree-capped redundancy graphs and modular penalties preserve support, whereas squared global cardinality penalties create all-to-all couplings. The supplement provides complete proofs, backend and optimization results, precision and checkpoint studies, the capability evidence behind Table~\ref{tab:capability}, and real-data and sampling details.

\section{Conclusion}

\method{} makes exact bounded-cone QUBO-QAOA objective-and-gradient queries a planned execution contract. Building on established causal-cone and adjoint identities, it profiles cone structure and induced-edge counts, jointly selects microbatches and checkpoints under a named workspace budget, and rejects requests that do not fit before local-amplitude and named-workspace allocation. Its outputs agree with an independent dense adjoint to $1.56\times10^{-13}$ worst relative error. The $p{=}2$ study extends beyond the reference's prespecified cap; an uncapped probe locates its allocation wall. Repeated-query results reduce training cost without changing the target. Global states, sampling, and large low-width contractions remain complementary routes.

\clearpage
\bibliography{references}

\end{document}